\documentclass[reprint,amsmath,amssymb,aps,floatfix,raggedbottom]{revtex4-2}
\usepackage{graphicx}
\usepackage{xcolor}
\usepackage{booktabs}
\usepackage{colortbl}
\usepackage{calc}
\usepackage{natbib}
\usepackage{hyperref}
\usepackage{tikz}
\usetikzlibrary{positioning,arrows.meta,calc,shapes.geometric,decorations.pathreplacing}
\usepackage[normalem]{ulem}

\definecolor{cFound}{HTML}{5B6573}    
\definecolor{cTeacher}{HTML}{38548C}  
\definecolor{cStudent}{HTML}{46B3AC}  
\definecolor{cMoire}{HTML}{79C9A8}    

\newcommand{\UWMSE}{Department of Materials Science and Engineering, University of Washington, Seattle, Washington 98195, USA\looseness=-1}
\newcommand{\PNNLfirst}{National Security Directorate, Pacific Northwest National Laboratory, Richland, Washington 99354, USA\looseness=-1}
\newcommand{\PNNLsecond}{Integrated Discovery Sciences Directorate, Pacific Northwest National Laboratory, Richland, Washington 99354, USA\looseness=-1}
\newcommand{\UWPHYS}{Department of Physics, University of Washington, Seattle, Washington 98195, USA\looseness=-1}

\begin{document}

\title{Distilling universal machine-learning potentials for moir\'e lattices across one million atoms}

\author{Thomas Huang$^{1}$}    
\thanks{These two authors contributed equally.}
\author{Yueyao Fan$^{1}$}  
\thanks{These two authors contributed equally.}
\author{Kaichen Xie$^{1}$}      
\author{Bolun Li$^{1}$}   
\author{Kaijie Yang$^{1}$}   
\author{Jenna A. Bilbrey$^{2}$} 
\author{Eric Bylaska$^{3}$}   
\author{Peter V. Sushko$^{3}$}  
\email{peter.sushko@pnnl.gov} 
\author{Di Xiao$^{1,4}$}
\email{dixiao@uw.edu} 
\author{Ting Cao$^{1}$}         
\email{tingcao@uw.edu} 
\affiliation{$^1$\UWMSE} 
\affiliation{$^2$\PNNLfirst}
\affiliation{$^3$\PNNLsecond}
\affiliation{$^4$\UWPHYS}

\begin{abstract}

Atomic reconstruction reshapes moir\'e materials across multiple scales, from local structure and polarization textures to global electronic topology, yet direct \textit{ab initio} modeling becomes prohibitive for large superstructures such as marginal-twist-angle moir\'es and moir\'e-of-moir\'es.
We develop Moir\'eMLIP by fine-tuning a universal atomistic model on the density functional theory labeled Moir\'e Kaleidoscope dataset, which spans transition metal dichalcogenide compositions, symmetries, stackings, and twist angles.
Moir\'eMLIP reproduces \textit{ab initio} reconstruction with force errors of 6--8\,meV/\AA{} and transfers to smaller twist angles and unseen structures.
Knowledge distillation yields Moir\'eMLIP-mini, which retains this accuracy while extending single-GPU inference to one million atoms.
Applied to an alternate-twist MoTe$_2$ trilayer, it reveals a polarization network spanning tens of nanometers arising from large moir\'e-cell distortions sharply localized along the moir\'e-of-moir\'e domain walls.
These results overcome key accuracy, transferability, and scaling bottlenecks of existing atomistic models and enable predictive simulations across emergent moir\'e length scales.
\end{abstract}

\maketitle

\section{Introduction}

Moir\'e materials define a materials design space with high-dimensional structural degrees of freedom, including twist angle, chemical composition, interlayer stacking, and layer number \cite{Geim2013nature,Kennes2021natphys,Andrei2021natrevmater,Mak2022natnano}, that are largely absent in conventional bulk crystals.
These degrees of freedom span distinct length scales, from chemistry and bonding at the unit-cell scale that determine intralayer elasticity to stacking textures at the moir\'e scale governed by interlayer stacking energetics.
Their coupling gives rise to engineerable moir\'e structures that reconstruct through the competition between these energetic contributions \cite{Naik2018prl,Yoo2019natmater,Cazeaux2020arma,Rosenberger2020acsnano,Weston2020natnano,Enaldiev2020prl,Halbertal2021natcomm,VanWinkle2023natcomm}.
Such reconstruction plays a decisive role in moir\'e properties, giving rise, for example, to flat Chern bands in twisted MoTe$_2$ \cite{Wang2024prl,Zhang2024natcomm} that underpin fractional Chern insulating (FCI) phases \cite{Cai2023nature,Zeng2023nature,Park2023nature,Cao2026arcmp}, and is therefore a prerequisite for understanding and predicting structure--property relationships. 
Therefore, systematic exploration of the high-dimensional moir\'e design space demands accurate and rapid structural relaxation across thousands of distinct configurations, from moir\'e cells to moir\'e-of-moir\'e superstructures containing $10^4$--$10^6$ atoms, which are often computationally prohibitive even within the restricted class of twisted transition metal dichalcogenide (TMD) homo- and hetero-structures. 

Existing approaches, however, face major bottlenecks at these scales.
Density functional theory (DFT) sets the accuracy standard for moir\'e structures, but its computational cost grows approximately cubically with system size when using plane-wave basis sets, making these calculations impractical.
Continuum elastic theory can efficiently model large moir\'e cells, but does not explicitly resolve all atomic-scale details, and some implementations consider only in-plane relaxation without explicitly including out-of-plane corrugation \cite{Carr2018prb,Zhu2020prb,NakatsujiKoshino2023prx,Nakatsuji2025commat}.
Classical force fields provide atomic coordinates, but their fixed functional forms, often at the few-parameter level, describe the stacking-registry energetics and the resulting reconstruction with insufficient accuracy \cite{Georgaras2026natcomm,Tran2026arxiv}. 
Machine-learned interatomic potentials (MLIPs)~\cite{Behler2007prl} trained from scratch on individual twisted systems can recover near-DFT forces \cite{Mao2024commphys,Fan2026pnas,Magorrian2025prm,Siddiqui2025twodmater,Shaidu2025npj,Liu2025npj,Bu2026jmps}, but their development typically requires labeling many configurations with DFT to cover the chemical compositions and structural environments of interest  \cite{Liu2025npj}.
This approach therefore does not scale to the combinatorial complexity of the high-dimensional moir\'e design space because of the large amount of DFT-labeled training data required.

\begin{figure*}[t!]
    \centering

    \begin{minipage}[c]{0.64\textwidth}
    \newsavebox{\wfbox}
    \sbox{\wfbox}{\resizebox{\linewidth}{!}{
    \begin{tikzpicture}[
        >=Latex,
        proc/.style={
            rounded corners=2pt,
            draw,
            line width=0.9pt,
            fill=cFound!12,
            align=center,
            inner sep=4pt,
            minimum height=12mm,
            minimum width=22mm,
            font={\sffamily\small}
        },
        model/.style={
            rounded corners=3pt,
            draw,
            line width=1pt,
            align=center,
            inner sep=4pt,
            font={\sffamily\small}
        },
        db/.style={
            cylinder,
            shape border rotate=90,
            aspect=0.28,
            draw,
            line width=0.9pt,
            fill=cMoire!22,
            align=center,
            inner sep=4pt,
            minimum width=16mm,
            minimum height=14mm,
            font={\sffamily\small}
        },
        flow/.style={
            -{Latex[length=2.6mm,width=2.0mm]},
            line width=1.1pt
        }
    ]
        \def\W{11.4}
        \def\yb{-3.4}

        \node[anchor=west, inner sep=0pt] (moimg) at (0,0)
            {\includegraphics[width=19mm]{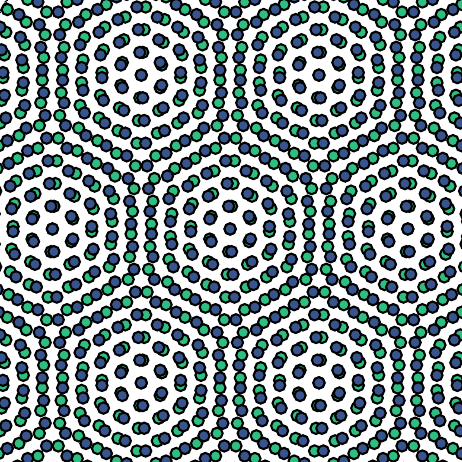}};

        \node[
            font={\sffamily\small},
            align=center,
            below=1.5mm of moimg
        ] (clab) {rigid moir\'e stacks};

        \node[proc, right=7mm of moimg] (amlip) {
            uMLIP\\
            relaxation and\\
            sampling
        };

        \node[proc, right=7mm of amlip] (alabel) {
            ab initio\\
            calculations\\
            label $E,\,F,\,\sigma$
        };

        \node[db, anchor=east] (adb) at (\W,0) {
            \textbf{Moir\'e}\\
            \textbf{Kaleidoscope}\\
            dataset
        };

        \draw[flow] (moimg.east) -- (amlip.west);
        \draw[flow] (amlip) -- (alabel);
        \draw[flow] (alabel) -- (adb.west);

        \node[
            proc,
            anchor=west,
            minimum width=20mm
        ] (bfound) at (0,\yb) {uMLIP};

        \node[
            model,
            fill=cTeacher!30,
            draw=cTeacher,
            minimum width=26mm,
            minimum height=16mm,
            right=14mm of bfound
        ] (bteacher) {
            \textbf{Moir\'eMLIP}\\
            teacher\\
            25.4M params
        };

        \node[
            db,
            right=7mm of bteacher
        ] (bsurr) {
            surrogate\\
            moir\'e\\
            database
        };

        \node[
            model,
            fill=cStudent!48,
            draw=cTeacher,
            minimum width=24mm,
            minimum height=16mm,
            anchor=east
        ] (bstud) at (\W,\yb) {
            \textbf{Moir\'eMLIP-mini}\\
            student\\
            2.3M params
        };

        \coordinate (ftmid)
            at ($(bfound.east)!0.5!(bteacher.west)$);

        \draw[flow]
            (bfound.east)
            -- node[
                font={\sffamily\small},
                below,
                inner sep=2pt
            ] {fine-tune}
            (bteacher.west);

        \draw[flow] (bteacher.east) -- (bsurr.west);
        \draw[flow] (bsurr.east) -- (bstud.west);

        \coordinate (brl)
            at ($(bteacher.south west)+(0,-0.22)$);
        \coordinate (brr)
            at (brl -| bstud.east);

        \draw[
            decorate,
            decoration={
                brace,
                amplitude=4pt,
                mirror,
                raise=1pt
            }
        ] (brl) -- (brr);

        \node[
            font={\sffamily\small},
            anchor=north,
            yshift=-2.5mm
        ] at ($(brl)!0.5!(brr)$)
            {knowledge distillation};

        \draw[
            flow,
            cMoire!80!black,
            line width=1pt
        ] (adb.south) -- ++(0,-0.8) -| (ftmid);

        \path (-0.85,0);
    \end{tikzpicture}
    }}

    \begin{tikzpicture}[remember picture]
        \node[
            anchor=south west,
            inner sep=0pt
        ] (wf) {\usebox{\wfbox}};

        \begin{scope}[x={(wf.south east)}, y={(wf.north west)}]
            \node[
                font={\sffamily\bfseries},
                anchor=north west,
                inner sep=1pt
            ] at (0.00,1.00) {(a)};

            \node[
                font={\sffamily\bfseries},
                anchor=north west,
                inner sep=1pt
            ] at (0.00,0.42) {(b)};

            \coordinate (dbhook) at (1.00,0.82);
        \end{scope}
    \end{tikzpicture}
    \end{minipage}\hfill
    \begin{minipage}[c]{0.35\textwidth}
    \raggedleft
    \begin{tikzpicture}[remember picture]
        \node[
            anchor=south west,
            inner sep=0pt
        ] (cimg) {
            \includegraphics[width=0.9\linewidth]{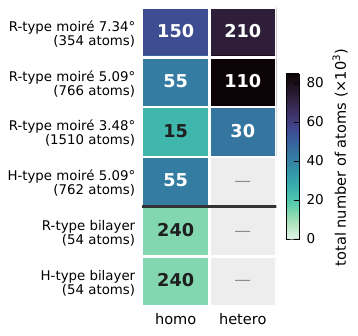}
        };

        \node[
            font={\sffamily\bfseries},
            anchor=south west,
            inner sep=1pt
        ] at ([xshift=-7pt]cimg.north west) {(c)};
    \end{tikzpicture}
    \end{minipage}

    \begin{tikzpicture}[remember picture, overlay]
        \coordinate (cwlo) at
            ($([xshift=-3pt]cimg.south west)!0.36!
              ([xshift=-3pt]cimg.north west)$);

        \coordinate (cwhi) at
            ($([xshift=-3pt]cimg.south west)!0.97!
              ([xshift=-3pt]cimg.north west)$);

        \path let
            \p1=(dbhook),
            \p2=(cwlo),
            \p3=(cwhi)
        in
            \pgfextra{
                \pgfmathsetmacro{\tmpasp}{(\y1-\y2)/(\y3-\y2)}
                \global\edef\braceasp{\tmpasp}
            };

        \draw[
            decorate,
            decoration={
                brace,
                amplitude=17pt,
                aspect=\braceasp
            },
            line width=1pt,
            black!60
        ] (cwlo) -- (cwhi);
    \end{tikzpicture}

    \caption{
        Workflow and dataset overview.
        (a)~Construction of the Moir\'e Kaleidoscope dataset.
        (b)~Fine-tuning uMLIP and knowledge distillation.
        (c)~Dataset coverage across composition and twist angle.
        Each cell is a category.
        Cell colors correspond to the total atom count summed over all supercell configurations in a category.
        The number inside a cell is the count of supercell configurations in that category.
    }
    \label{fig:workflow}
\end{figure*}

Universal MLIPs (uMLIPs) promise the missing transferability, providing a single foundation model across much of the periodic table \cite{Batatia2025jcp,Fu2025icml,Rhodes2025arxiv,Kavanagh2026arxiv}.
In the case of moir\'e superlattices, however, uMLIPs face two challenges. 
The first is that the moir\'e structures lie outside the training domain of existing uMLIPs. 
Their training sets consist primarily of bulk crystals with no more than $\sim 100$ atoms per supercell \cite{BarrosoLuque2024arxiv} and, therefore, contain no interlayer sliding configurations, moir\'e supercells, or long-period strain fields generated by reconstruction.
Their validity for moir\'e reconstruction is therefore unclear, and no systematic benchmark exists, particularly given that uMLIP accuracy is known to degrade upon transition from bulk to lower-dimensional structures \cite{Benedini2025aifs}.
Indeed, as we show below, the ten uMLIPs we benchmark exhibit widely varying zero-shot force errors of 18--86\,meV\,\AA$^{-1}$. 
The second obstacle, and perhaps the more fundamental one, is that model size and the corresponding inference cost tend to increase with accuracy requirements.
In the moir\'e context, the most accurate uMLIP can exceed typical GPU memory limits of hundreds of gigabytes at only $\sim 10^3$ atoms, leaving important structures such as marginal-twist-angle bilayers, multilayer stacks, and larger emergent superstructures computationally inaccessible \cite{Fan2026pnas,Nakatsuji2025commat,Adak2024natrevmater}. 

Here we overcome these challenges through fine-tuning and knowledge distillation. 
First, we assemble the Moir\'e Kaleidoscope dataset, comprising more than one thousand DFT-labeled TMD moir\'e structures.
Fine-tuning a uMLIP on this dataset yields Moir\'eMLIP, reducing the force error by more than fivefold while retaining transferability across the moir\'e design space.
The post-training learning curve follows a power-law scaling behavior, showing strong few-instance performance and continuous improvement with increasing training-data volume.
We then distill Moir\'eMLIP into Moir\'eMLIP-mini, a compact 2.3-million-parameter force model \cite{Wang2025prm,Gardner2025arxiv,Matsumura2026springer,Oh2026jcim} that achieves comparable accuracy with a $\sim28\times$ smaller memory footprint.
On a single GPU, Moir\'eMLIP-mini scales to one million atoms, more than 20 times the system-size ceiling of its teacher and 500 times that of the most accurate uMLIP in our benchmark.
Applying Moir\'eMLIP-mini to moir\'e-of-moir\'e MoTe$_2$ trilayers with different twist angles enables fully atomistic relaxation at previously inaccessible length scales \cite{Zhang2026advmat,Craig2024natmater,Hoke2026natmater,Uri2023nature,Devakul2023sciadv,Lin2022prb,Zhu2020prb,NakatsujiKoshino2023prx,Nakatsuji2025commat,Meng2023prb} and reveals an emergent polarization network arising from reconstruction-induced electric polarization.

\section{Moir\'e Kaleidoscope Dataset}
\label{subsec:dataset_generation}

To enable systematic benchmarking and post-training of uMLIPs for moir\'e reconstruction, we assemble the Moir\'e Kaleidoscope dataset.
Unlike general-purpose uMLIP training corpora, which do not systematically sample the structural degrees of freedom associated with moir\'e reconstruction, Moir\'e Kaleidoscope provides a common DFT-labeled reference spanning TMD compositions, stackings, and twist angles.
This common reference enables direct comparison of zero-shot uMLIPs, controlled fine-tuning, and quantitative analysis of post-training scaling.
The initial release of the Moir\'e Kaleidoscope (V1.0) dataset provides \textit{ab initio} labels for TMD moir\'e structures (Fig.~\ref{fig:workflow}). 
This set includes structures formed from five hexagonal TMD monolayers (MoS$_2$, MoSe$_2$, MoTe$_2$, WS$_2$, and WSe$_2$), comprising five homobilayer and ten heterobilayer combinations.
We sample R-type moir\'e structures at twist angles of 7.34$^\circ$, 5.09$^\circ$, and 3.48$^\circ$, as well as H-type moir\'e structures for homobilayers at 5.09$^\circ$ [Fig.~\ref{fig:workflow}(c)].
We also include untwisted homobilayer supercells with R- and H-type high-symmetry stackings.

To construct the dataset and labels, each category of structures passes through a three-stage pipeline. 
First, a pretrained uMLIP with vdW-D2 correction is used to relax the initial structure to residual forces below 0.1\,eV\,\AA$^{-1}$, followed by molecular dynamics (MD) simulations at 1{,}000\,K using the same uMLIP \cite{Rhodes2025arxiv,Grimme2006}.
Details of the structural sampling protocol are provided in Appendix~\ref{app:sampling}.
Structural configurations are then sampled from the MD trajectories and labeled with DFT-PBE energies, forces, and stresses (Appendix~\ref{app:dft}).
The DFT labels exclude the vdW-D2 contribution, thereby decoupling the long-range dispersion interaction from the DFT-PBE potential energy surface (PES) learned by the MLIP.
This design provides a common PBE target for consistent benchmarking across different uMLIPs and allows a chosen dispersion correction, such as vdW-D2 or vdW-D3, to be added separately in subsequent MLIP calculations \cite{Sauer2025prm}.

The dataset contains 1{,}105 labeled configurations, split into 710 for training and 395 for testing, with the 3.48$^\circ$ cells held out entirely for out-of-distribution testing (Appendix Table~\ref{tab:dataset_construction}).

\section{Training and Validation of Moir\'eMLIP}
\label{sec:training_validation}

As we show below, the tested uMLIPs do not reach the \textit{ab initio} accuracy required to reliably reproduce key features of moir\'e lattice reconstruction, motivating the incorporation of moir\'e-specific training data.
We therefore fine-tune orb-v3-conservative \cite{Rhodes2025arxiv}, selected for its favorable inference cost, on the Moir\'e Kaleidoscope dataset.
We refer to the resulting fine-tuned model as Moir\'eMLIP.

\begin{figure}[!b]
    \centering
    \includegraphics[width=0.75\columnwidth]{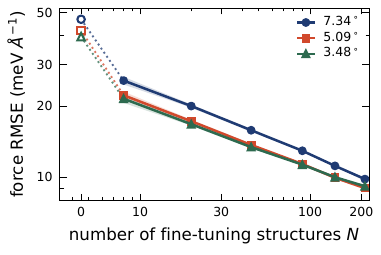}
    \caption{Fine-tuning learning curves across twist angles. 
    Force RMSE as a function of the number of fine-tuning structures $N$. 
    Models are fine-tuned using only the 7.34$^\circ$ structures and evaluated at 7.34$^\circ$ as well as on the held-out 5.09$^\circ$ and 3.48$^\circ$ structures, demonstrating transferability across twist angle. 
    Filled markers and shaded bands show the mean and one standard deviation over five random seeds during training, respectively; open markers at $N=0$ denote the zero-shot uMLIP baseline for each test set.}
    \label{fig:scaling}
\end{figure}

Our fine-tuning results demonstrate excellent few-instance performance and post-training scaling laws for continuous improvement of the model upon increasing the training data volume. 
We fine-tune on a growing number ($N$) of \(7.34^\circ\) moir\'e structures and evaluate the resulting models on \(7.34^\circ\) (hold-out set), \(5.09^\circ\), and \(3.48^\circ\) moir\'e structures.
The force root-mean-square error (RMSE) on the \(7.34^\circ\) structures in Fig.~\ref{fig:scaling} has a sharp drop from \(N=0\) to \(N=8\), showing that fine-tuning with only a handful of moir\'e configurations reduces the force RMSE by approximately a factor of two.
As $N$ increases further, the error continues to decrease, revealing a systematic post-training scaling behavior (Fig.~\ref{fig:scaling}).
Between $N=8$ and $N=210$, the RMSE follows a single power law, $\varepsilon_F \propto N^{-\alpha}$ with $\alpha = 0.29$, and shows no sign of saturation up to $N=210$.
The curves evaluated on the hold-out set of \(5.09^\circ\) and \(3.48^\circ\) moir\'e structures closely follow the \(7.34^\circ\) curve, showing that the few-instance model improvement and the subsequent scaling behavior are transferrable across twist angle.

We then perform a full-scale training of Moir\'eMLIP using the complete training set. 
The force RMSE on the hold-out set decreases from 40.8 to 7.5\,meV\,\AA$^{-1}$, and the energy RMSE decreases from 5.11 to 0.73\,meV\,atom$^{-1}$. 
Remarkably, the force RMSE stays below 8\,meV\,\AA$^{-1}$ at every tested twist angle and stacking category (Table~\ref{tab:table1}). 
We benchmark ten uMLIPs zero-shot on the same test set in Appendix~\ref{app:zeroshot}, with force RMSE ranging from 18 to 86\,meV\,\AA$^{-1}$. 
The cutoff radius is kept fixed at 6~\AA, identical to that of the pretrained orb-v3 model, indicating that the improved accuracy does not arise from extending the interaction range.
The complete fine-tuning parameters and protocol are provided in Table~\ref{tab:training_hyperparameters}.

\begin{table*}[t]
  \centering
  \caption{Force and energy RMSE of orb-v3, Moir\'eMLIP, and Moir\'eMLIP-mini across six moir\'e test subsets (Table~\ref{tab:dataset_construction}). 
  Cell shading indicates RMSE on separate logarithmic color scales for force (6.2--43.9~meV\,\AA$^{-1}$) and energy (0.25--6.54~meV\,atom$^{-1}$), each shared across all models and test subsets.}
  \label{tab:table1}
  \setlength{\tabcolsep}{4pt}
  \begin{tabular}{lcccccc}
    \toprule
    & \multicolumn{3}{c}{Force RMSE (meV\,\AA$^{-1}$)} & \multicolumn{3}{c}{Energy RMSE (meV\,atom$^{-1}$)} \\
    \cmidrule(lr){2-4}\cmidrule(lr){5-7}
    Subset & \makebox[\widthof{Moir\'eMLIP-mini}][c]{orb-v3} & \makebox[\widthof{Moir\'eMLIP-mini}][c]{Moir\'eMLIP} & Moir\'eMLIP-mini & \makebox[\widthof{Moir\'eMLIP-mini}][c]{orb-v3} & \makebox[\widthof{Moir\'eMLIP-mini}][c]{Moir\'eMLIP} & Moir\'eMLIP-mini \\
    \midrule
    R $7.34^\circ$ & \cellcolor[HTML]{7CB0D3}43.9 & \cellcolor[HTML]{F0F6FA}7.8 & \cellcolor[HTML]{E3EEF6}9.4 & \cellcolor[HTML]{86B6D6}5.15 & \cellcolor[HTML]{DBE9F3}0.61 & \cellcolor[HTML]{EBF3F8}0.41 \\
    R $5.09^\circ$ & \cellcolor[HTML]{7FB2D4}41.9 & \cellcolor[HTML]{F7FAFC}7.0 & \cellcolor[HTML]{EAF2F8}8.5 & \cellcolor[HTML]{87B7D7}4.96 & \cellcolor[HTML]{DFEBF4}0.56 & \cellcolor[HTML]{F8FBFD}0.30 \\
    R $3.48^\circ$ & \cellcolor[HTML]{83B4D5}39.4 & \cellcolor[HTML]{F0F6FA}7.7 & \cellcolor[HTML]{E7F0F7}8.9 & \cellcolor[HTML]{8CB9D8}4.42 & \cellcolor[HTML]{EFF6FA}0.37 & \cellcolor[HTML]{FFFFFF}0.25 \\
    H $5.09^\circ$ & \cellcolor[HTML]{83B4D5}39.8 & \cellcolor[HTML]{F2F7FB}7.5 & \cellcolor[HTML]{E4EFF6}9.3 & \cellcolor[HTML]{7CB0D3}6.54 & \cellcolor[HTML]{D2E4F0}0.76 & \cellcolor[HTML]{E3EEF6}0.50 \\
    R high-sym. & \cellcolor[HTML]{97C0DC}29.4 & \cellcolor[HTML]{FFFFFF}6.2 & \cellcolor[HTML]{FFFFFF}6.2 & \cellcolor[HTML]{88B7D7}4.87 & \cellcolor[HTML]{D7E7F2}0.68 & \cellcolor[HTML]{DFEBF4}0.56 \\
    H high-sym. & \cellcolor[HTML]{9BC3DD}27.6 & \cellcolor[HTML]{FDFEFE}6.4 & \cellcolor[HTML]{FBFCFE}6.6 & \cellcolor[HTML]{84B5D6}5.35 & \cellcolor[HTML]{C3DBEB}1.12 & \cellcolor[HTML]{C4DBEB}1.09 \\
    \midrule
    \textbf{Overall} & \cellcolor[HTML]{81B3D5}40.8 & \cellcolor[HTML]{F2F7FB}7.5 & \cellcolor[HTML]{E7F0F7}8.9 & \cellcolor[HTML]{86B6D6}5.11 & \cellcolor[HTML]{D4E5F1}0.73 & \cellcolor[HTML]{DBE9F3}0.61 \\
    \bottomrule
  \end{tabular}

  \vspace{2.5mm}
  \definecolor{cbmax}{HTML}{7CB0D3}
  \begin{tikzpicture}
    \shade[left color=white, right color=cbmax] (0,0) rectangle (4.2,0.18);
    \draw[gray!60, line width=0.3pt] (0,0) rectangle (4.2,0.18);
    \node[anchor=east, font=\footnotesize] at (-0.15,0.09) {lower RMSE};
    \node[anchor=west, font=\footnotesize] at (4.35,0.09) {higher};
  \end{tikzpicture}
\end{table*}

\begin{figure*}[t]
    \centering
    \includegraphics[width=\textwidth]{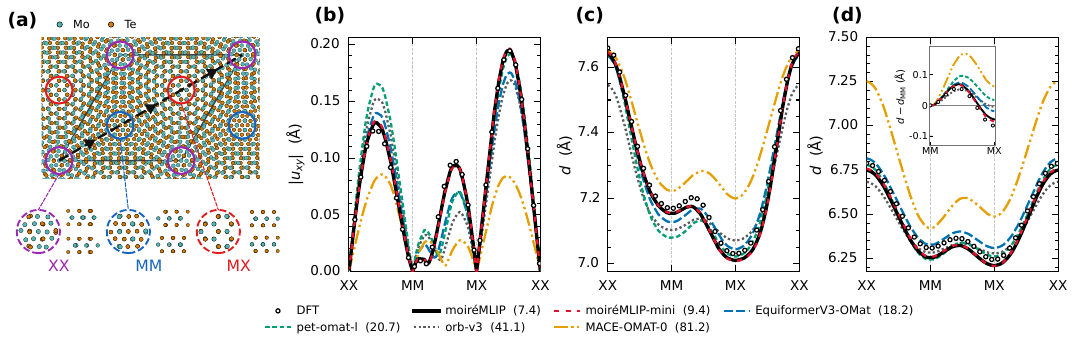}
    \caption{Reconstruction of H-type bilayers along a line cut through the stacking registries. 
    (a)~Top view of H-type MoTe$_2$ at 4.41$^\circ$ in its DFT-relaxed geometry, the same structure measured in (b) and (c). 
    The gray parallelogram is a moir\'e cell and the dashed diagonal with arrows is the line cut of the XX-MM-MX-XX path used in (b) and (c). 
    Circles mark the three stacking registries with zoomed-in top and side views. 
    (b),(c)~In-plane displacement magnitude $|u_{xy}|$ of metal atoms (b) and interlayer distance $d$ (c), with the DFT-relaxed reference drawn as open circles. 
    (d)~Interlayer distance along the XX-MM-MX-XX path in H-type MoS$_2$ at 4.41$^\circ$, using the same convention as (c). 
    The inset shows the MM to MX segment, with each curve plotting the interlayer distance relative to its value at MM, i.e., ($d-d_{\mathrm{MM}}$).
    In (b-d), different curves correspond to results obtained from Moir\'eMLIP, Moir\'eMLIP-mini, and four zero-shot uMLIPs (EquiformerV3-OMat, pet-omat-l, orb-v3, and MACE-OMAT-0). 
    Line styles follow the legend, in which the number after each model's name denotes its force RMSE in meV\,\AA$^{-1}$.}
    \label{fig:accuracy}
\end{figure*}

The accuracy of uMLIPs has improved rapidly in recent years and is often evaluated in terms of global force RMSE.
While this is an important metric, a small global force RMSE alone does not guarantee correct moir\'e reconstruction.
The longer-period nature of moir\'e structures can amplify local reconstruction errors even when the average force RMSE over many atoms in a moir\'e supercell remains unchanged or decreases, which can mask local force errors and significantly affect the reconstructed structure.
To validate the models directly on moir\'e reconstruction, we relax two H-type homobilayer moir\'es, MoTe$_2$ and MoS$_2$, both at \(4.41^\circ\) [Fig.~\ref{fig:accuracy}(a)]. 
These systems approach the largest sizes accessible to the selected uMLIPs benchmarked in Fig.~\ref{fig:accuracy}, i.e., EquiformerV3-OMat, pet-omat-l, orb-v3, and MACE-OMAT-0, which were chosen either for their strong performance on the Moir\'e Kaleidoscope benchmark (Appendix~\ref{app:zeroshot}) or for their relatively small memory footprint.
The DFT and all uMLIP relaxations are converged to a maximum residual force below 5\,meV\,\AA$^{-1}$.

Remarkably, Moir\'eMLIP closely reproduces the DFT reconstruction profiles along the XX-MM-MX-XX diagonal line, where XX (chalcogen on chalcogen), MM (metal on metal), and MX (metal on chalcogen) are points in the moir\'e cell with three-fold rotational axes [Fig.~\ref{fig:accuracy}(a)].
In contrast, the pretrained uMLIPs fail to capture key reconstruction features at zero shot.
The principal failure of the pretrained uMLIPs is an underestimation of the contrast in interlayer distance and intralayer displacement between distinct configurations, likely arising from their compressed variations in relative energies between different local stacking configurations.
The pretrained uMLIPs also shift the peak and dip positions, indicating misplaced energetic features.
These deficiencies may reflect inadequate coverage of moir\'e-relevant stacking configurations in the training data.

In MoTe$_2$ moir\'es, specifically, the pretrained uMLIPs, in comparison to Moir\'eMLIP, misrepresent the relative magnitudes or positions of the peaks in the in-plane displacement profile [Fig.~\ref{fig:accuracy}(b)]. 
The pretrained uMLIPs generally yield a more similar strain between XX-MM and between MX-XX, by overestimating the peak between the former and underestimating that between the latter.
Between MM-MX, the two smaller peaks and the minimum in strain are all shifted towards MX, and the pronounced difference between the peak heights is substantially underestimated.
The pretrained uMLIPs also underestimate the MM–MX interlayer-distance contrast, reducing the DFT value of 0.140~\AA{} to 0.025–0.084~\AA{} [Fig.~\ref{fig:accuracy}(c) and Table~\ref{tab:reconstruction_metrics}].
Similarly, in MoS$_2$ moirés, the peak-position shifts persist, and the pretrained models severely underestimate, or even qualitatively mispredict, the difference between the MM and MX interlayer distances, reducing the DFT contrast of 0.067~\AA{} to between $-0.062$ and $0.019$~\AA{} [Fig.~\ref{fig:accuracy}(d) and Table~\ref{tab:reconstruction_metrics}].
These results highlight the need to benchmark reconstruction accuracy directly, rather than rely solely on global force accuracy, when evaluating MLIPs for moir\'e systems and potentially beyond.

\section{Distillation of Moir\'eMLIP}
\label{sec:distill}

\begin{figure*}[!t]
    \centering
    \includegraphics[width=0.8\textwidth]{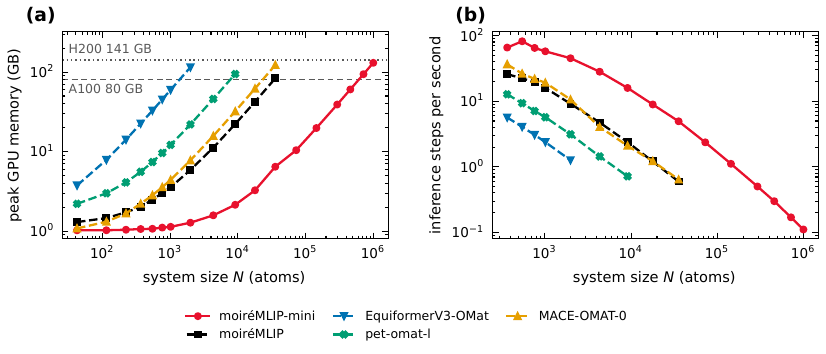}
    \caption{Inference cost of Moir\'eMLIP-mini, Moir\'eMLIP, and three selected uMLIPs in logarithmic scale. 
    (a)~Peak GPU memory usage during inference (the lower the better) and (b)~inference steps completed per second (the higher the better), on a single NVIDIA H200, versus system size $N$ (number of atoms) of MoTe$_2$ homobilayer moir\'e supercells, with full FP32 precision.  
    Moir\'eMLIP-mini (solid) predicts forces directly through a force prediction head.
    The other four models (dashed) are conservative, with forces differentiated from the energy. 
    Each curve ends at the largest benchmarked size at its memory ceiling. The horizontal lines in (a) mark 141\,GB of an H200 (dotted) and the 80\,GB of an A100 (dashed), two types of GPUs used in our work.}
    \label{fig:cost}
\end{figure*}

Moir\'eMLIP accurately predicts lattice reconstruction across the moir\'e design space. 
However, Moir\'eMLIP retains the size and architecture of the pretrained model, and is therefore memory limited, precluding its general application to very small-angle, multilayer, and moir\'e-of-moir\'e systems (Fig.~\ref{fig:cost}(a)). 
For these structures, the central challenge for model designs is the vast scale of the superstructure during the model inference and simultaneously maintaining sufficient accuracy. 

We address this challenge through teacher-student knowledge distillation \cite{Wang2025prm,Gardner2025arxiv,Matsumura2026springer,Oh2026jcim}, allowing for task-specific compression of Moir\'eMLIP to a compact student model, Moir\'eMLIP-mini.
Moir\'eMLIP-mini accurately reproduces the force output and lattice reconstruction calculated using Moir\'eMLIP, while supporting systems $\sim28\times$ larger and inference $\sim7\times$ faster. 
As a result, Moir\'eMLIP-mini crosses an otherwise inaccessible length scale of uMLIPs, supporting single-GPU inference for systems with one million atoms in under 10 seconds per inference step.

On the materials side, the compression is enabled by focusing on moir\'e TMDs only, which include a small set of compounds.
As the monolayers are in the same 1H crystal phase, the structural diversity arises from stacking, twisting, and 3D deformation. 
The student may therefore prioritize these chemical and structural degrees of freedom that govern reconstruction, and thereby preserve the accuracy of the teacher in these areas, without spanning the full chemical and structural spaces.

On the modeling side, the large gains in inference speed and the ability to handle large system sizes over Moir\'eMLIP arise from the two principles, i.e., the direct-force and deep-narrow design, of Moir\'eMLIP-mini.
This direct-force formulation focuses on the relaxation task, which determines the important low-energy electronic structure of the reconstructed moir\'e superlattices. 
Having accurate lattice reconstruction requires reliable atomic forces near the minimum of the potential energy surface, but does not require strict energy conservation at every ionic step \cite{Bigi2025icml,Rhodes2025arxiv}.
As the forces are predicted in the forward pass, the activations of each message-passing layer can be released once that layer has been evaluated, leaving peak memory governed mainly by the latent width \cite{Park2024jctc}.
Our architecture sweep further shows that accuracy is governed primarily by the total parameter count. 
The model capacity can therefore be distributed over deeper message-passing layers with a narrower latent representation, preserving the accuracy required for reconstruction prediction while reducing the memory cost.

We distill Moir\'eMLIP into Moir\'eMLIP-mini by training the student force model from scratch on two separate datasets: the 710 \textit{ab initio}-labeled structures used to fine-tune Moir\'eMLIP and 21,833 additional molecular-dynamics and relaxation snapshots labeled by Moir\'eMLIP. 
The latter dataset is generated rapidly by Moir\'eMLIP without expensive \textit{ab initio} calculations.
Moir\'eMLIP-mini (2.3 million parameters) is 9\% of Moir\'eMLIP's size, yet its force RMSE is only 20\% larger than that of Moir\'eMLIP (Table~\ref{tab:table1}). 

Moir\'eMLIP-mini brings nearly identical moir\'e reconstruction patterns compared with Moir\'eMLIP results [Fig.~\ref{fig:accuracy}(b)--(d)], and enables research of large systems previously inaccessible to Moir\'eMLIP. 
On a single NVIDIA H200 GPU, Moir\'eMLIP-mini supports inference for structures of 1 million atoms, increasing the accessible system size to $\sim28\times$ and accelerating inference speed by $\sim7\times$ (at $10^4$ atoms) compared with Moir\'eMLIP (Fig.~\ref{fig:cost}).
Moir\'eMLIP-mini also reaches systems 500 times larger than EquiformerV3,~\cite{Liao2026arxiv} one of the most accurate uMLIPs in our benchmark, and runs 36 times faster at EquiformerV3's $2\times10^3$-atom ceiling on H200 (Fig.~\ref{fig:cost}). 

\section{Polarization textures of Moir\'e-of-moir\'e structures}
\label{subsec:trilayer_demo}

\begin{figure*}[t]
    \centering
    \includegraphics[width=0.8\textwidth]{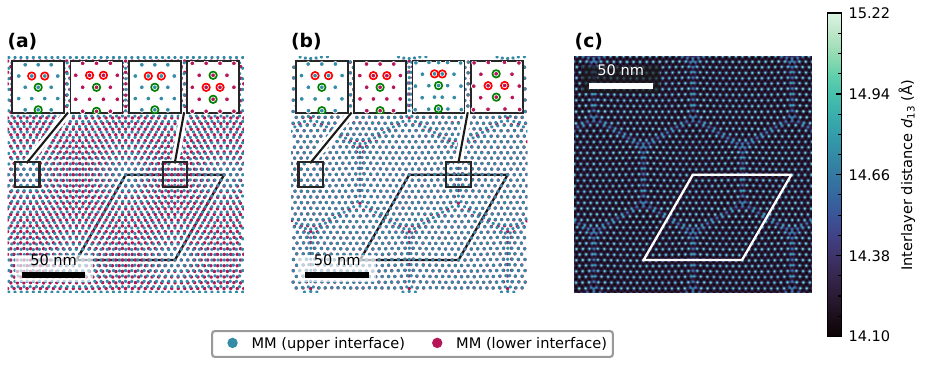}
    \caption{Moir\'e patterns and interlayer distance of the alternate-twist MoTe$_2$ trilayer moir\'e-of-moir\'e ($\theta_{12}=-4.15^\circ$, $\theta_{23}=+3.89^\circ$).
    (a) MM-site positions at the top (layers 1 and 2, blue) and bottom (layers 2 and 3, red) moir\'e interfaces before relaxation.
    (b) MM-site positions after relaxation.
    (c) Color map of the relaxed Mo--Mo distance between the outer layers, $d_{13}=z_1-z_3$.
    The parallelograms outline one moir\'e-of-moir\'e cell.
    Insets in (a) and (b) show the MM sites at each interface separately for the boxed regions.
    The left and right inset pairs in (a) and (b) show the regions that form an MM-aligned domain and a domain wall, respectively, before (a) and after (b) relaxation.
    Red and green circles mark selected MM-site pairs used to compare spacings perpendicular and parallel to the domain wall, respectively.}
    \label{fig:trilayer}
\end{figure*}

Moir\'e-of-moir\'e structures arise from the interference of two moir\'e patterns with slightly different periodicities, generating a longer-wavelength superstructure beyond the individual moir\'e length scales \cite{Zhang2026advmat,Craig2024natmater,Hoke2026natmater,Uri2023nature,Devakul2023sciadv,Lin2022prb,Zhu2020prb,NakatsujiKoshino2023prx,Nakatsuji2025commat,Meng2023prb}.
Structural relaxation can further reorganize this interference pattern into extended locally commensurate domains separated by domain walls.
In TMDs, the resulting reconstruction can generate electric polarization through ferroelectric and piezoelectric responses that underlie the strong moir\'e potential \cite{Wang2022natnano,Weston2022natnano,Sangers2025small}.
Although the reconstruction-induced polarization is well established in bilayer moir\'es \cite{Wang2022natnano,Weston2022natnano,Li2025natcomm,Enaldiev2020prl,Magorrian2021prb,Zhang2024natcomm,Zhang2025npjqm}, its microscopic organization in moir\'e-of-moir\'e structures remains largely unexplored.
Moir\'eMLIP-mini enables us to investigate this regime by relaxing an alternate-twist MoTe$_2$ trilayer with $\theta_{12}=-4.15^\circ$ and $\theta_{23}=+3.89^\circ$, where both twist angles are close to those at which FCI phases have been observed in twisted bilayer MoTe$_2$.
The interference between the two moir\'e patterns produces a 77-nm superstructure containing 439{,}473 atoms.

We find that lattice reconstruction gives rise to large strain and distortions at the moiré-cell scale, sharply localized along the moiré-of-moiré domain walls [Fig.~\ref{fig:trilayer}]. 
This behavior is distinct from the atomic-scale reconstruction within individual moiré cells, where the atomic displacements remain small. 
Instead, these small local displacements accumulate coherently across moiré cells, producing a pronounced rearrangement at the larger moiré-cell scale. 
This reconstruction is characterized by the relative rearrangement of the MM-stacking regions of moiré cells at the two moiré interfaces [Fig.~\ref{fig:trilayer}(a),(b)].

\begin{figure*}[!t]
    \centering
        \includegraphics[width=0.8\textwidth]{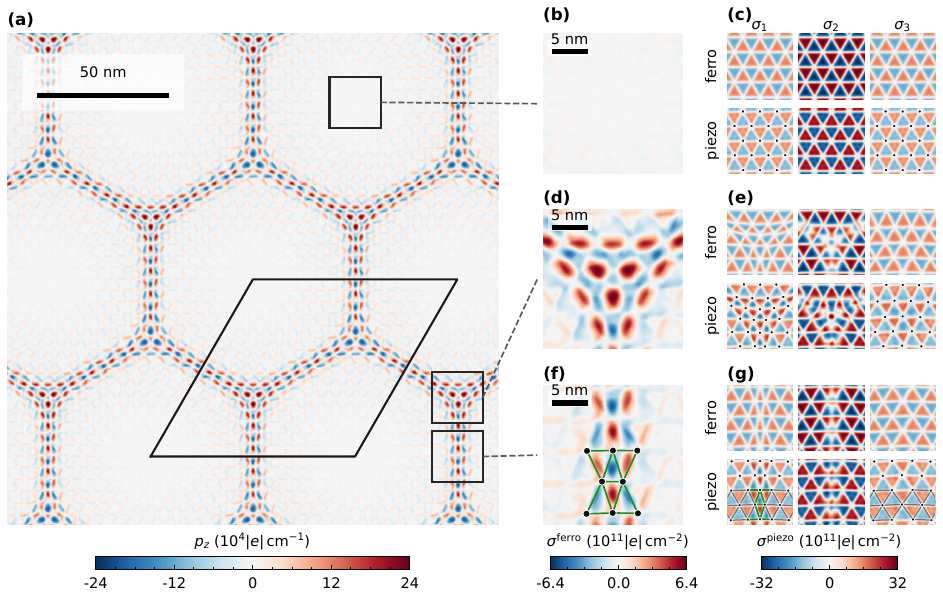}
    \caption{Surface charge of the alternate-twist MoTe$_2$ trilayer moir\'e-of-moir\'e of Fig.~\ref{fig:trilayer}, evaluated from the relaxed geometry.
    (a) Out-of-plane polarization $p_z$ over the moir\'e-of-moir\'e structure.
    The boxes mark the regions shown in (b), (d), and (f), and the parallelogram outlines one moir\'e-of-moir\'e cell.
    (b),(d),(f) Enlarged maps of the corresponding out-of-plane polarization $p_z$.
    (c),(e),(g) Ferroelectric and piezoelectric bound charge densities $\sigma_1$, $\sigma_2$, and $\sigma_3$ on the first, second, and third layers, respectively.
    Black dots in the $\sigma_1$ and $\sigma_3$ piezoelectric charge maps mark the MM stacking sites of the top and bottom moir\'e interfaces, respectively.
    Gray hexagons in (g) highlight regions where the top moir\'e interface contains one more MM site than the bottom moir\'e interface along a horizontal cut.
    The green mesh in the $\sigma_1$ piezoelectric charge map outlines the associated extra moir\'e domains.
    The same MM-site markers and green mesh are superimposed in (f), illustrating how the sign alternation and spatial geometry of the piezoelectric charge in these domains are reflected in $p_z$.}
    \label{fig:charge}
\end{figure*}

In the rigid structure, the MM stacking sites of the upper and lower interfaces form uniform arrays with slightly different periodicities [Fig.~\ref{fig:trilayer}(a)].
After relaxation, MM stacking sites from the two interfaces locally align to form extended and commensurate domains [Fig.~\ref{fig:trilayer}(b)].
However, because the two interfaces contain different numbers of MM stacking sites within one moir\'e-of-moir\'e cell, complete MM-site alignment is impossible.
The resulting mismatch is accommodated by interconnected networks of higher-energy domains and domain walls, consistent with prior work on alternate-twist WSe$_2$ trilayers~\cite{Nakatsuji2025commat}.
The higher-energy domains shrink during relaxation, accompanied by an approximately $C_3$-symmetric compression and dilation of the MM-site pattern at the upper and lower interfaces, respectively [Fig.~\ref{fig:trilayer}(b)].
At the domain walls, the MM-site deformation becomes strongly anisotropic [Fig.~\ref{fig:trilayer}(b)].
The red markers in the inset of Fig.~\ref{fig:trilayer}(b) track the MM-site spacing of the upper interface perpendicular to the wall, which decreases by 42.3\%, from 50.0\,\AA{} in the MM-aligned domain [inset of Fig.~\ref{fig:trilayer}(a)] to 28.9\,\AA{}.
At the lower interface, the corresponding spacing instead increases by 22.4\%, from 50.1\,\AA{} in the MM-aligned domain to 61.3\,\AA{} at the wall.
In contrast, the spacing parallel to the wall, indicated by the green circles in the inset of Fig.~\ref{fig:trilayer}(b), remains nearly unchanged.
The domain-wall reconstruction is therefore characterized by localized and predominantly one-dimensional compression and dilation of moir\'e unit cells at the upper and lower interfaces, respectively, perpendicular to the wall.

The full atomistic relaxation also resolves the out-of-plane reconstruction, manifested in the spatial variation of the outer-layer separation $d_{13}=z_1-z_3$ [Fig.~\ref{fig:trilayer}(c)].
Within the MM-aligned domains, $d_{13}$ exhibits extended regions with sharp, regularly ordered maxima, reflecting the local alignment of the MM stacking sites at the two interfaces.
By contrast, at the narrow domain walls and their intersections, reduced oscillations of $d_{13}$ lead to a blurry pattern in Fig.~\ref{fig:trilayer}(c), because the MM sites of the two interfaces avoid each other in the in-plane directions. 

The relaxed atomistic structure allows us to calculate the out-of-plane electric polarization of the alternate-twist MoTe$_2$ based on the atomic displacement and interlayer lateral shift. 
We find that the interconnected network of higher-energy domains and domain walls [Fig.~\ref{fig:trilayer}(b)] is directly reflected in the polarization texture [Fig.~\ref{fig:charge}(a)].
Fig.~\ref{fig:charge}(b),(d),(f) show that the MM-aligned low-energy domains are nearly nonpolar, and out-of-plane electric polarization $p_z$ develops unique patterns where the two MM-site patterns from the two interfaces depart from local alignment.
In the higher-energy domain [Fig.~\ref{fig:charge}(d)], the approximately $C_3$-symmetric compression and dilation of the MM-site pattern at the upper and lower interfaces are accompanied by a similarly symmetric $p_z$ texture.
In comparison, at the domain wall [Fig.~\ref{fig:charge}(f)], the strongly anisotropic deformation of the MM-site pattern directly translates to a stripe-like $p_z$ texture.

The microscopic origin of this correspondence is revealed by the layer-resolved ferroelectric and piezoelectric bound charges in Fig.~\ref{fig:charge}(c),(e),(g), calculated as described in Appendix~\ref{app:polarization} \cite{Zhang2024natcomm,Zhang2025npjqm}.
The piezoelectric charge substantially exceeds the ferroelectric charge in magnitude, consistent with the bilayer moir\'e case.
Within the MM-aligned domains, the two moir\'e interfaces are locally registered and approximately related by horizontal mirror symmetry, leading to substantial cancellation of their contributions to $p_z$ despite finite bound charge on each layer.
Along a horizontal cut across the regions enclosed by the gray hexagons in Fig.~\ref{fig:charge}(g), the top interface contains one more MM site than the bottom interface.
The extra moir\'e domains associated with this site, outlined by the green mesh in $\sigma_1^{\mathrm{piezo}}$, carry alternating signs of piezoelectric charge and introduce an additional positive--negative--positive sequence, or its sign-reversed counterpart, along this horizontal cut [Fig.~\ref{fig:charge}(g)].
This extra sequence is not matched by the bottom-layer charge distribution and therefore gives rise to a localized $p_z$ sequence with the same sign alternation [Fig.~\ref{fig:charge}(f)].
The same MM-site markers and green mesh superimposed in Fig.~\ref{fig:charge}(f) show that the $p_z$ lobes follow the spatial arrangement and shapes of the piezoelectric charge regions in these extra domains, including the locations of sign reversals.
Strong reconstruction locks the regions on either side of the wall into near MM alignment, confining the interfacial mismatch to a width comparable to that of a single triangular domain at the bottom moir\'e interface.
Thus, the single excess MM site along this cut corresponds to one complete sign-alternation sequence in $p_z$, with the polarization returning to nearly zero in the MM-aligned regions on either side.
Across the supercell, the moir\'e-of-moir\'e domain structure
organizes the local $p_z$ textures into a polarization network
[Fig.~\ref{fig:charge}(a)].

These reconstruction-induced polarization textures could be
investigated using local experimental probes
\cite{Wang2022natnano,Weston2022natnano,Li2025natcomm,Sangers2025small},
including Kelvin probe force microscopy (KPFM) to map the
associated spatial variations in surface potential.
The resulting electrostatic landscape may influence carrier
localization and enable polarization-defined confinement in
moir\'e-of-moir\'e structures.

\section{Discussion}

Our fine-tuning and distillation framework overcomes the accuracy and scale limitations of uMLIPs for moir\'e structural problems.
We first construct the Moir\'e Kaleidoscope dataset to benchmark uMLIPs and fine-tune orb-v3. 
The resulting Moir\'eMLIP model recovers near-\textit{ab initio} forces and reconstruction across the TMD moir\'e design space.
We then use Moir\'eMLIP to generate a large surrogate moir\'e database and distill its predictive accuracy into Moir\'eMLIP-mini, extending single-GPU inference to million-atom systems while retaining the reconstruction accuracy of Moir\'eMLIP.
This workflow pushes MLIP-based transferable atomistic modeling beyond conventional moir\'e superlattices into the previously inaccessible moir\'e-of-moir\'es regime.
Fully atomistic relaxation reveals a polarization network in the alternate-twist MoTe$_2$ trilayer.
The locally commensurate domains and domain walls host distinct polarization textures arising from reconstruction-induced electric polarization.
More generally, because our distillation workflow requires only inferences from a trained model, it can be applied to generic uMLIPs or fine-tuned models to construct compact models for a target materials space and task at previously inaccessible scales.

\begin{acknowledgments}
This work is primarily supported by the Computational Materials Sciences Program funded by the U.S. Department of Energy, Office of Science, Basic Energy Sciences, Materials Sciences, and Engineering Division, PNNL FWP 83557.
Y.F. was supported by the U.S.-Japan University Partnership for Workforce Advancement and Research and Development in Semiconductors (UPWARDS).
This research used resources of the National Energy Research Scientific Computing Center, a DOE Office of Science User Facility supported by the Office of Science of the U.S. Department of Energy under Contract No. DE-AC02-05CH11231 using NERSC award BES-ERCAP0032546 and BES-ERCAP0033256.
This work was also facilitated through the use of advanced computational, storage, and networking infrastructure provided by the Hyak supercomputer system and funded by the University of Washington Molecular Engineering Materials Center at the University of Washington (DMR-2308979).
\end{acknowledgments}

\newpage

\appendix

\begin{center}
    \textbf{APPENDIX}
\end{center}

\section{\textit{Ab initio} calculations}
\label{app:dft}

\textit{Ab initio} calculations were performed using VASP 6.4.1 \cite{Kresse1996prb} within the projector augmented-wave (PAW) formalism \cite{Blochl1994prb,Kresse1999prb} and the PBE exchange-correlation functional \cite{Perdew1996prl}.
The PBE version 54 PAW potentials were used, with Mo\_pv and W\_sv for Mo and W, respectively.
Spin-orbit coupling and Hubbard-$U$ corrections were not included.

All structures in the Moir\'e Kaleidoscope dataset were constructed using experimental lattice constants. 
We used $a=3.16$~\AA\ for MoS$_2$ and WS$_2$~\cite{Villars2023MoS2,McMurdie1986,Villars2023WS2,Mandal1997}, $3.29$~\AA\ for MoSe$_2$ and WSe$_2$~\cite{Villars2023MoSe2,Villars2023WSe2,Mandal1997}, and $3.49$~\AA\ for MoTe$_2$~\cite{Agarwal1972}. 
Table~\ref{tab:dataset_construction} summarizes the training and test sets.

\section{Structural sampling}
\label{app:sampling}

Structural configurations for the Moir\'e Kaleidoscope dataset were generated by relaxation followed by finite-temperature molecular dynamics.
The initial moir\'e structures were first relaxed using a pretrained uMLIP with the Grimme D2 dispersion correction until the residual forces were below 0.1\,eV\,\AA$^{-1}$.
Starting from the relaxed structures, molecular dynamics was performed using Langevin dynamics in the canonical (NVT) ensemble at 1{,}000\,K.
Initial velocities were sampled from a Maxwell--Boltzmann distribution at 1{,}000\,K.
A timestep of 2\,fs and a friction coefficient of 0.01\,fs$^{-1}$ were used for all MD trajectories.
The $7.34^\circ$ structures were propagated for 200 steps (0.4\,ps), whereas the $5.09^\circ$ and $3.48^\circ$ structures were propagated for 100 steps (0.2\,ps).

Structural configurations were sampled from the MD trajectories for subsequent DFT labeling, with 21, 11 and 3 configurations  selected per moir\'e cell for the $7.34^\circ$, $5.09^\circ$, and $3.48^\circ$ structures, respectively.
Additional relaxation-trajectory configurations were included for the $7.34^\circ$ structures (Table~\ref{tab:dataset_construction}).
The resulting configurations were evaluated by single-point DFT calculations to obtain reference energies, forces, and stresses, as described in Appendix~\ref{app:dft}.

\begin{table}[tbp]
\caption{Composition and descriptive statistics of the Moir\'e Kaleidoscope dataset. Each subset is labeled by stacking (R or H) and twist angle; the two high-symmetry rows are untwisted reference cells at the R-stacked (MM, MX, XM) and H-stacked (MX, XX, MM) registries. The 7.34$^\circ$ row includes 45 relaxation-trajectory frames (30 train, 15 test); all other entries are molecular-dynamics snapshots.}
\label{tab:dataset_construction}
\begin{ruledtabular}
\begin{tabular}{lrrr}
Subset & Mean atom count & Train & Test \\
\hline
R $7.34^\circ$ & 354     & 240 & 120 \\
R $5.09^\circ$ & 766     & 105 & 60  \\
R $3.48^\circ$ & 1{,}510 & 0   & 45  \\
H $5.09^\circ$ & 762     & 35  & 20  \\
R high-sym.    & 54      & 165 & 75  \\
H high-sym.    & 54      & 165 & 75  \\
\hline
Total          &         & 710 & 395 \\
\end{tabular}
\end{ruledtabular}
\end{table}

\section{Structural relaxation}
\label{app:relaxation}

All MLIP-based structural relaxations reported in this work were performed using the FIRE optimizer \cite{Bitzek2006prl} implemented in the Atomic Simulation Environment (ASE), with the Grimme D2 dispersion correction described in Appendix~\ref{app:d2}.
To generate configurations for DFT labeling in the construction of the Moir\'e Kaleidoscope dataset, we used a pretrained uMLIP with the vdW-D2 correction implemented in \texttt{torch-dftd}.
When using Moir\'eMLIP models, atomic positions were relaxed without cell optimization, because the strain applied to construct commensurate structures is negligibly small. 
Space-group symmetry was preserved during relaxation using the symmetry constraints implemented in ASE to accelerate the calculations.
All relaxations were performed in full FP32 precision. For moir\'e of moir\'es, all atoms are relaxed to a maximum residual force below 3 meV/\AA\ with the cell dimensions fixed.
A uniform strain $<$ 0.01\% is applied to construct the commensurate superlattice.

\section{GPU-accelerated dispersion correction}
\label{app:d2}

The vdW-D2 dispersion correction~\cite{Grimme2006} was added to the MLIP as a separate contribution using \texttt{torch-dftd}~\cite{Takamoto2022natcomm}, a GPU implementation of vdW-D2/D3. 
The periodic neighbor-list construction in \texttt{torch-dftd}, however, remains CPU-based and becomes a computational bottleneck for large moir\'e supercells.
We therefore implemented a CUDA-based periodic neighbor list using \texttt{torch\_cluster}.
This implementation accelerates the vdW-D2 calculation by approximately sevenfold for a 100,000-atom bilayer moir\'e supercell.

\section{Zero-shot benchmark of uMLIPs}
\label{app:zeroshot}

Table~\ref{tab:zeroshot} benchmarks ten uMLIPs on the 395-structure Moir\'e Kaleidoscope test set. 
Six models were trained on OMat24: EquiformerV3~\cite{Liao2026arxiv}, PET~\cite{Bigi2026mlst}, eSEN~\cite{Fu2025icml}, GRACE~\cite{Lysogorskiy2026npj,Bochkarev2024prx}, orb-v3~\cite{Rhodes2025arxiv}, and MACE~\cite{Batatia2022neurips,Batatia2025jcp}. EquFlashV2~\cite{Lee2025icml}, NequIP~\cite{Kavanagh2026arxiv,Tan2026digitaldiscovery}, and Allegro~\cite{Musaelian2023natcomm} were trained on the combined OMat24+MPtrj+sAlex (OAM) dataset. 
For the multi-fidelity SevenNet-Omni-i12~\cite{Kim2026natcomm,Park2024jctc}, we report the OMat24-fidelity head.

The zero-shot force RMSE ranges from 18.2 to 86.3\,meV\,\AA$^{-1}$. 
EquiformerV3 achieves the lowest error of 18.2\,meV\,\AA$^{-1}$, followed by PET and eSEN at 20.7 and 21.2\,meV\,\AA$^{-1}$, respectively. 
orb-v3 gives a zero-shot error of 40.8\,meV\,\AA$^{-1}$. 

Despite the relatively low force errors of the best-performing uMLIPs, they do not reliably recover the \textit{ab initio} moir\'e reconstruction, as discussed in Sec.~\ref{sec:training_validation}. 
Fine-tuning orb-v3 on the Moir\'e Kaleidoscope dataset reduces the force RMSE to 7.5\,meV\,\AA$^{-1}$ and captures the moir\'e reconstruction.

\begin{table}[h]
\centering
\small
\caption{Zero-shot force RMSE of ten uMLIPs on the Moir\'e Kaleidoscope test set. All models are evaluated in full FP32 precision.}
\label{tab:zeroshot}
\begin{tabular}{llc}
\toprule
Architecture & Pretrained weights & \shortstack{Force RMSE\\(meV\,\AA$^{-1}$)} \\
\midrule
EquiformerV3 & EquiformerV3-OMat   & 18.2 \\
PET          & pet-omat-l          & 20.7 \\
eSEN         & eSEN-30M-OMat       & 21.2 \\
GRACE        & GRACE-2L-OMAT-large-base    & 39.2 \\
ORB          & orb-v3-conservative-inf-omat & 40.8 \\
EquFlashV2   & EquFlashV2-45M-OAM  & 57.4 \\
SevenNet     & SevenNet-Omni-i12   & 60.6 \\
NequIP       & NequIP-OAM-XL       & 60.9 \\
MACE         & MACE-OMAT-0         & 81.2 \\
Allegro      & Allegro-OAM-L       & 86.3 \\
\bottomrule
\end{tabular}
\end{table}

\section{Calculation of ferroelectric and piezoelectric polarization}
\label{app:polarization}

The bound surface charges in Fig.~\ref{fig:charge} were evaluated from the relaxed geometry using the continuum description of moir\'e polarization developed in Refs.~\cite{Zhang2024natcomm,Zhang2025npjqm}.
The in-plane displacement field $\mathbf{u}_\ell(\mathbf{r})$ was obtained from the Mo positions of the relaxed structure relative to the rigid stack, while $z_\ell(\mathbf{r})$ denotes the relaxed Mo-plane height.
The local interlayer distance of interface $p$ formed by the upper layer $\ell$ and lower layer $\ell'$ is $d_p(\mathbf{r})=|z_\ell(\mathbf{r})-z_{\ell'}(\mathbf{r})|$.

The ferroelectric charge originates from interlayer charge transfer that depends on the local stacking registry \cite{Zhang2025npjqm}.
For each interface, the registry is described by the stacking function
\begin{equation}
S_p(\mathbf{r})=
\sum_{j=1}^{3}\sin\left[
\mathbf{g}^{(p)}_j\cdot\mathbf{r}
+\mathbf{G}^{(\ell)}_j\cdot\mathbf{u}_\ell(\mathbf{r})
-\mathbf{G}^{(\ell')}_j\cdot\mathbf{u}_{\ell'}(\mathbf{r})
\right],
\label{eq:stacking}
\end{equation}
where $\mathbf{G}^{(\ell)}_j$ and $\mathbf{G}^{(\ell')}_j$ are the first-shell reciprocal lattice vectors of the two layers, including their respective rotations and uniform strains.
The moir\'e reciprocal lattice vectors of interface $p$ are defined as
$\mathbf{g}^{(p)}_j=\mathbf{G}^{(\ell')}_j-\mathbf{G}^{(\ell)}_j$.
The ferroelectric surface charge of interface $p$ is
\begin{equation}
\sigma^{\mathrm{ferro}}_p(\mathbf{r})
=-\epsilon_0\,
\frac{\tilde v_f(d_p^*)}{d_p^*}\,
S_p(\mathbf{r}),
\label{eq:ferro}
\end{equation}
which is placed on layer $\ell$ with the opposite sign on layer $\ell'$.
Here $\tilde v_f(d)$ is the ferroelectric potential amplitude of Ref.~\cite{Zhang2025npjqm}, and the prefactor follows from the Poisson relation between the ferroelectric charge and potential given there.
The effective interlayer distance is $d_p^*=f_\sigma\bar d_p$.
Here $\bar d_p$ is the average interlayer distance weighted by the square of the local ferroelectric charge density evaluated using $d_p(\mathbf{r})$, following Eq.~(28) of Ref.~\cite{Zhang2025npjqm}.
In the trilayer, the two interfaces are treated as independent bilayers and their charges on the middle layer are superposed.
This approximation neglects coupling between charge transfer at the upper and lower interfaces.

The piezoelectric charge of each layer is evaluated independently from the gradient of its strain field \cite{Zhang2024natcomm,Zhang2025npjqm},
\begin{equation}
\sigma^{\mathrm{piezo}}_\ell(\mathbf{r})
=p_\sigma\,\tilde e_{11}
\left[
2\partial_x u_{xy}
+\partial_y\left(u_{xx}-u_{yy}\right)
\right]_\ell,
\label{eq:piezo}
\end{equation}
where the strain components and derivatives are expressed in the crystallographic frame of layer $\ell$, and $\tilde e_{11}$ is the piezoelectric coefficient of monolayer MoTe$_2$.
The strain tensor is defined as
$u_{ij,\ell}=(\partial_i u_{j,\ell}+\partial_j u_{i,\ell})/2$.
Screening is included through the spatially uniform factor $p_\sigma=0.2021$ from Ref.~\cite{Zhang2025npjqm}.
The resulting $\sigma_\ell^{\mathrm{piezo}}$ is an effective screened charge density.
The same screening factor obtained for the bilayer is applied to all three layers.

We evaluate the local out-of-plane areal dipole density about the supercell midplane as
\begin{equation}
p_z(\mathbf{r})=
\sum_{\ell=1}^{3}
\left[z_\ell(\mathbf{r})-z_{\mathrm{ref}}\right]
\sigma_\ell(\mathbf{r}),
\label{eq:pz}
\end{equation}
where $\sigma_\ell=\sigma_\ell^{\mathrm{ferro}}+\sigma_\ell^{\mathrm{piezo}}$ and the reference height $z_{\mathrm{ref}}$ is taken at the center of the supercell along $z$.

We used $\tilde v_f(d)=p_0+p_1(d-d_0)+p_2(d-d_0)^2$ with the coefficients, $f_\sigma$, and the piezoelectric coefficient $\tilde e_{11}=2.68\times10^{-10}\,\mathrm{C\,m^{-1}}$ taken from Table~1 of Ref.~\cite{Zhang2025npjqm}.
The displacement fields, interlayer distances, stacking functions, and strain gradients were computed from our relaxed structures.

\clearpage
\onecolumngrid

\begin{table}[!ht]
  \centering
  \caption{Registry-resolved and global reconstruction metrics for H-type
  MoTe$_2$ and MoS$_2$ bilayers at $4.41^\circ$.
  $\Delta d_{\mathrm{MM-MX}}=d_{\mathrm{MM}}-d_{\mathrm{MX}}$ is the
  interlayer-distance difference between the MM and MX registries.
  $\sigma_d$ and $\sigma_{|u_{xy}|}$ are the spatial standard deviations
  of the interlayer distance and in-plane displacement magnitude over the
  full moir\'e supercell, respectively. All quantities are in \AA.}
  \label{tab:reconstruction_metrics}
  \setlength{\tabcolsep}{7pt}

  \begin{tabular}{lcccccc}
    \toprule
    & \multicolumn{3}{c}{MoTe$_2$}
    & \multicolumn{3}{c}{MoS$_2$} \\
    \cmidrule(lr){2-4}
    \cmidrule(lr){5-7}
    Model
    & $\Delta d_{\mathrm{MM-MX}}$
    & $\sigma_d$
    & $\sigma_{|u_{xy}|}$
    & $\Delta d_{\mathrm{MM-MX}}$
    & $\sigma_d$
    & $\sigma_{|u_{xy}|}$ \\
    \midrule
    DFT
    & 0.140 & 0.147 & 0.047
    & 0.067 & 0.138 & 0.015 \\
    Moir\'eMLIP
    & 0.144 & 0.151 & 0.047
    & 0.046 & 0.138 & 0.018 \\
    Moir\'eMLIP-mini
    & 0.147 & 0.152 & 0.047
    & 0.053 & 0.142 & 0.017 \\
    \midrule
    EquiformerV3
    & 0.084 & 0.141 & 0.044
    & 0.019 & 0.132 & 0.014 \\
    PET
    & 0.058 & 0.150 & 0.048
    & $-0.017$ & 0.139 & 0.018 \\
    orb-v3
    & 0.033 & 0.115 & 0.045
    & 0.005 & 0.106 & 0.017 \\
    MACE
    & 0.025 & 0.116 & 0.022
    & $-0.062$ & 0.211 & 0.007 \\
    \bottomrule
  \end{tabular}
\end{table}

\begin{table}[!ht]
\caption{Training hyperparameters for Moir\'eMLIP and Moir\'eMLIP-mini.}
\label{tab:training_hyperparameters}
\begin{ruledtabular}
\begin{tabular}{lcc}
& Moir\'eMLIP & Moir\'eMLIP-mini \\
\hline
Optimizer & AdamW & AdamW \\
Epochs & 1000 & 1000 \\
Batch size & 8 & 16 \\
Maximum learning rate & $5\times10^{-5}$ & $5\times10^{-4}$ \\
Minimum learning rate & $5\times10^{-6}$ & $1\times10^{-4}$ \\
Learning-rate schedule & cosine & cosine \\
Warmup epochs & 10 & 20 \\
Weight decay & $1\times10^{-6}$ & $1\times10^{-4}$ \\
Energy loss weight & 16 & 1 \\
Force loss weight & 1 & 8 \\
Stress loss weight & 0.25 & 0.25 \\
Equigrad loss weight & $1\times10^{-5}$ & — \\
Cutoff radius (\AA) & 6.0 & 6.0 \\
Maximum neighbors & 60 & 60 \\
\end{tabular}
\end{ruledtabular}
\end{table}

\clearpage
\twocolumngrid

\end{document}